\documentclass[aps,prl,twocolumn,superscriptaddress,amsmath]{revtex4-2}
\usepackage{graphicx}
\usepackage{hyperref}
\usepackage{mathrsfs}
\usepackage{bm}
\usepackage{color}
\usepackage{siunitx}
\usepackage[capitalize]{cleveref}
\usepackage{orcidlink}

\hypersetup{hypertex=true,
	colorlinks=true,
	anchorcolor=blue,
	linkcolor=blue,
    citecolor=blue,
	urlcolor=blue}
\begin{document}

\title{Anti-higher-order topological insulators}

\author{Cheng-Ming Miao\,\orcidlink{0000-0002-8095-8053}}
%\email[]{These authors contribute equally to this work.}
\affiliation{International Center for Quantum Materials, School of Physics, Peking University, Beijing 100871, China}

\author{Yu-Hao Wan\,\orcidlink{0009-0007-3049-8355}}
\email[]{wanyh@stu.pku.edu.cn}
\affiliation{International Center for Quantum Materials, School of Physics, Peking University, Beijing 100871, China}

\author{Ying-Tao Zhang\,\orcidlink{0000-0003-2783-1325}}
\affiliation{College of Physics, Hebei Normal University, Shijiazhuang 050024, China}

\author{Qing-Feng Sun\,\orcidlink{0000-0002-5512-9608}}
\email[]{sunqf@pku.edu.cn}
\affiliation{International Center for Quantum Materials, School of Physics, Peking University, Beijing 100871, China}
\affiliation{Hefei National Laboratory, Hefei 230088, China}

\begin{abstract}
Duality is a fundamental concept in physics that connects complementary opposites like particles and holes. Similarly, while topological states in topological insulators localize at boundaries, the existence and nature of their dual counterparts remain unexplored. Here, we introduce anti-topological states as the dual of topological states, exemplified by anti-higher-order topological insulators. Unlike higher-order topological insulators, where states localize at corners, anti-higher-order topological insulators host states along edges but absent at corners, realizing an inverted distribution of states. We demonstrate this phenomenon in a bilayer Chern insulator with opposite Chern numbers, where the band inversion surfaces enclose distinct high-symmetry points. The topological invariant distinguishing these phases is given by the topological charges enclosed by band inversion surfaces. This work establishes anti-topology as a new paradigm, opening a chapter in topological research.

\vspace{1em}

\end{abstract}

\maketitle

\emph{Introduction}.---Duality is a unifying principle in physics: it captures the idea that two apparently opposite descriptions in fact represent complementary aspects of the same underlying principle, often providing a more complete and unified understanding of physical phenomena \cite{savit_duality_1980,fruchart_dualities_2020,zhang_general_2023}. A classic example in condensed matter is particle–hole duality, where the absence of an electron acts as a positively charged quasiparticle \cite{nguyen_particlehole_2017}. Such a dual perspective provides a unifying framework for understanding a wide range of quantum phenomena. In recent years, the concept of topology has reshaped our understanding of quantum matter, culminating in the discovery of higher-order topological (HOT) insulators \cite{benalcazar_quantized_2017,schindler_higherorder_2018,serra-garcia_observation_2018,peterson_quantized_2018,xie_secondorder_2018,xie_visualization_2019,chen_direct_2019,zhang_secondorder_2019,sheng_twodimensional_2019,ren_engineering_2020,peterson_fractional_2020}. Unlike conventional topological insulators where gapless states appear along all boundaries \cite{qi_topological_2011,rhim_unified_2018}, HOT insulators extend the bulk–boundary correspondence by hosting boundary modes of reduced dimensionality. In general, a $d$-dimensional $n$-th order topological system hosts gapless modes on its $(d-n)$-dimensional edge states; for instance, a two-dimensional second-order topological insulator can exhibit zero-dimensional corner states \cite{miao_secondorder_2022,miao_engineering_2023,miao_general_2024,guo_secondorder_2024,liu_engineering_2024,liu_engineering_2025,miao_tunable_2025}. Yet from the viewpoint of duality, this picture remains incomplete. Just as particles have holes as their dual counterparts, one may ask whether there exists a converse of HOT insulator—namely, a $d$-dimensional insulator where topological states persist along most $(d-1)$-dimensional boundaries but are absent on the lower-dimensional boundaries. Identifying such anti-HOT phases would uncover the dual structure of topology and open a new avenue in the study of topological matter.

\begin{figure}
  \centering
\includegraphics[width=0.9\columnwidth,angle=0]{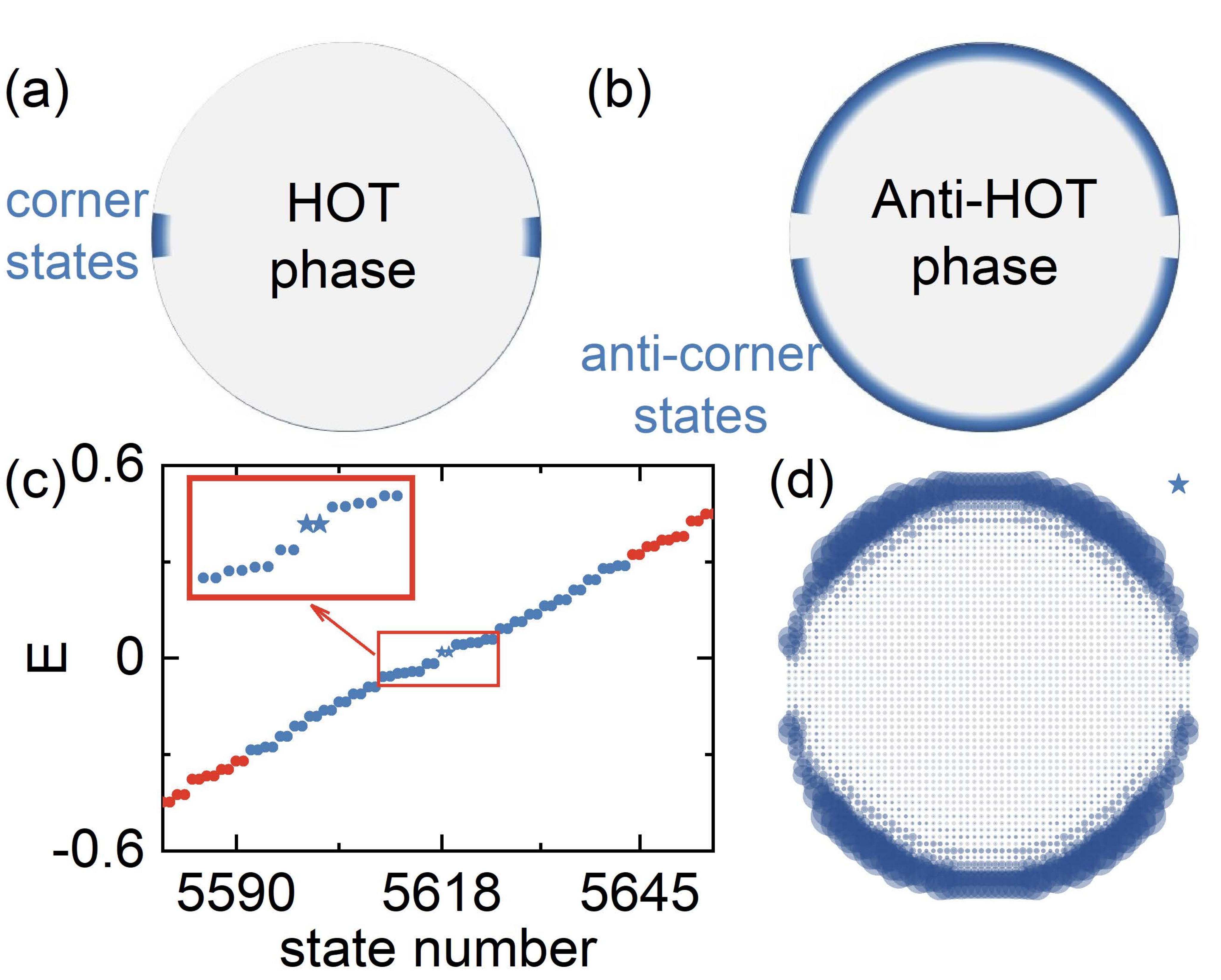}
  \caption{(a, b) Schematic of the HOT phase with localized corner states in a circular nanoflake (a), and the anti-HOT phase with anti-corner states extending along most boundaries but vanishing at specific corners (b). (c) Numerical energy levels of a finite-sized circular bilayer Chern insulator system, where anti-corner states are highlighted in blue symbols. The red rectangle indicates the near zero energy region, which is enlarged in the upper left inset. (d) Spatial distributions of representative anti-corner states  marked by the star symbols in (c). The gray dots represent discrete lattice sites and the areas of the blue circles are proportional to the local density of states. The parameters are set as $R=30, M_1=M_2=1,A_x^1=A_y^1=A_x^2=-A_y^2=1,B_x^1=B_y^1=1,B_x^2=1,B_y^2=-2,t_0=0.3$.}
  \label{fig1}
\end{figure}

In this Letter, we propose this missing dual counterpart, which we term the anti-HOT insulators. Their defining phenomenon is a spatially inverted distribution of topological states relative to HOT insulators: in two dimensions, while HOT insulators host localized corner states [Fig. \ref{fig1}(a)], the anti-HOT insulators instead exhibit anti-corner states—extended modes that appear along most boundaries but vanish at the corners [Fig. \ref{fig1}(b)]. We demonstrate this dual behavior in a bilayer lattice model with opposite Chern numbers, whose numerical spectra and spatial distributions clearly reveal the persistence of anti-corner states. Furthermore, altermagnetically coupled Chern insulators are predicted to provide a potential platform for realizing the anti-HOT phase. The topological invariant distinguishing these phases is given by the topological charges enclosed by the band inversion surfaces (BIS). Our work introduces anti-topology as a new theoretical framework, opening a new frontier in the exploration of topological states and guiding future investigations in this emerging field.

\emph{Toy model}.---
We first consider a bilayer Chern insulator model with opposite Chern numbers. This setup can be realized in synthetic systems such as photonic or cold-atom platforms \cite{Liang_realization_2023,braun_realspace_2024,wang_hybrid_2023,lai_photonic_2025,mandal_photonic_2025}. As a minimal model, it serves to demonstrate anti-HOT phases. In the absence of interlayer coupling, the system hosts two counter-propagating first-order edge states. Each layer (labeled $\alpha=1, 2$) is described by the modified Qi–Wu–Zhang (QWZ) model \cite{qi_topological_2006} in momentum space $\mathbf{k}=\left(k_x,k_y\right)$: $H_\alpha\left(\mathbf{k}\right)=\mathbf{d}^\alpha\left(\mathbf{k}\right)\cdot\mathbf{\sigma}$, where $\mathbf{\sigma}=\left(\sigma_x,\sigma_y,\sigma_z\right)$ are the Pauli matrices acting on the spin subspace. The components of the $\mathbf{d}^\alpha$-vector for each layer can be given by:
\vspace{-0.5em}
\begin{align}
d_z^\alpha\left(\mathbf{k}\right)&=M_\alpha-2B_x^\alpha\left(1-\cos{k_x}\right)-2B_y^\alpha\left(1-\cos{k_y}\right),\nonumber \\ d_x^\alpha\left(\mathbf{k}\right)&=A_x^\alpha\sin{k_x},\qquad \quad \quad d_y^\alpha\left(\mathbf{k}\right)=A_y^\alpha\sin{k_y}.
\label{eq1}
\end{align}
Here, $M_\alpha$ is the mass parameter, while $A_{x/y}^\alpha$ and $B_{x/y}^\alpha$ denote the spin–orbit coupling and hopping strengths along the $x/y$ direction in layer $\alpha$, respectively. The $d_x^\alpha\left(\mathbf{k}\right)$ and $d_y^\alpha\left(\mathbf{k}\right)$ terms generate in-plane spin texture, whereas the $d_z^\alpha\left(\mathbf{k}\right)$ term controls the BIS \cite{wan_classification_2025}. The topology of each layer is characterized by the Chern number $C_\alpha$, which is calculated by integrating the Berry curvature over the Brillouin zone \cite{xiao_berry_2010}. The two layers are coupled via spin-independent nearest-neighbor hopping with strength $t_0$, giving the bilayer Hamiltonian: $\mathcal{H}\left(\mathbf{k}\right)=\left(\begin{matrix}H_1\left(\mathbf{k}\right)&t_0\sigma_0\\\left(t_0\sigma_0\right)^\ast&H_2\left(\mathbf{k}\right)\\\end{matrix}\right)$, where $\sigma_0$ is the identity matrix.
For the numerical demonstration, we consider a bilayer system in which the two layers are chosen to realize opposite Chern numbers ($C_1=+1$  and $C_2=-1$) \cite{zhao_tuning_2020}, ensuring the presence of counter-propagating edge states in the absence of interlayer coupling ($t_0=0$). Details of the square-lattice tight-binding Hamiltonian and the specific parameter choices are provided in Sec. S1 of the Supplemental Material \cite{supplemental}.

Next, we perform numerical simulations on a finite-size circular nanoflake. The geometry of the circular nanoflake is constructed on a square lattice plane centered at the origin $\left(0,0\right)$, where the position of each site is labeled by its coordinates $(p,q)$. The sites belonging to the circular nanoflake satisfy the condition $p^2+q^2<R^2$ with radius $R$.
We note that the circular shape only specifies the real-space boundary of the finite nanoflake and does not imply continuous rotational invariance of the lattice Hamiltonian. The system is defined on a square lattice, and the parameters used below further break the equivalence between the $x$ and $y$ directions. In particular, we take $A_x^2 \neq A_y^2$ and $B_x^2 \neq B_y^2$ in layer $2$, making the Hamiltonian explicitly direction dependent. Figure \ref{fig1}(c) shows the near zero-energy spectrum, highlighting a set of states (blue  symbols) whose spatial distributions we analyze below. For clarity, the red rectangle indicates the near zero energy region, which is enlarged in the upper left inset of Fig. \ref{fig1}(c). The star symbols mark the representative state selected for visualizing its spatial distribution. As shown in Fig.~\ref{fig1}(d), these states are localized along most boundaries but are strongly suppressed near specific boundary positions.
These states exhibit typical characteristic: they are localized along edges but vanish sharply along specific boundary directions, as shown in Fig. \ref{fig1}(d). We define these states as anti-corner states, which serve as the defining feature of the anti-HOT phase. Supplementary Video and Sec. S2 of the Supplemental Material \cite{supplemental} provide animated visualizations of the local density distributions for every blue-symbol state, confirming that the directional absence along specific boundaries is a robust property shared by all of these states. For completeness, Sec. S3 and Fig. S1 of the Supplemental Material \cite{supplemental} presents the total spatial distributions of all states appearing in the energy spectrum.
Together, these results demonstrate the persistence and directional character of anti-corner states, providing direct evidence for the existence of the anti-HOT phase in this bilayer Chern insulator model.

\begin{figure}
	\centering
	\includegraphics[width=\columnwidth,angle=0]{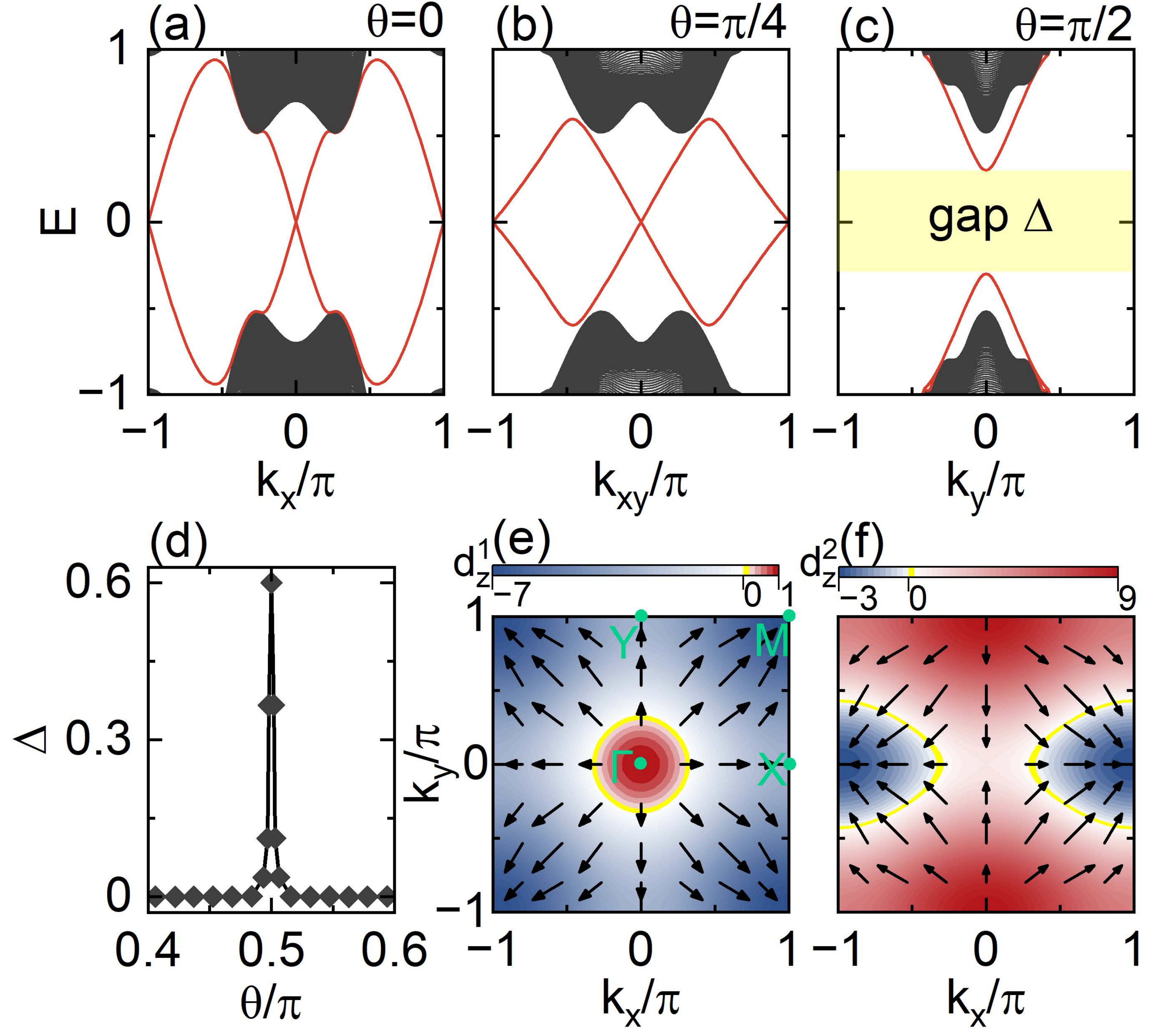}
	\caption{(a-c) Band structures of nanoribbons with periodic boundary conditions along different edge directions. The edge angles are $\theta=0$ in (a), $\theta=\pi/4$ in (b), and $\theta=\pi/2$ in (c). Red and black lines correspond to edge states and bulk states, respectively. (d) The energy gap $\Delta$ as a function of edge angles $\theta$. (e, f) Spin-texture maps in the Brillouin zone for layer $1$ and layer $2$, respectively. The black arrows represent the components of ($d_x^\alpha, d_y^\alpha$), while the color represents the $d_z^\alpha$ component (with the yellow contour marking $d_z^\alpha=0$). Four green dots in (e) correspond to four high-symmetry points. The nanoribbon width is set to $L=300$, and all other parameters are the same as those in Fig. \ref{fig1}.}
	\label{fig2}
\end{figure}

\emph{The origin of anti-HOT phase}.---To elucidate the underlying mechanism, we examine nanoribbons with periodic boundary conditions along different directions in Fig. \ref{fig2}(a-c), where $\theta$ denotes the edge angle relative to the $x$-axis (see Fig. S2 and Sec. S4 of the Supplemental Material \cite{supplemental}).
Notably, an energy gap $\Delta \neq 0$ opens only for $\theta=\pi/2$ [Fig. \ref{fig2}(c)], while edge states (red lines) along other orientations remain gapless [Figs. \ref{fig2}(a) and \ref{fig2}(b)]. A systematic survey of edge orientations reveals that the gap magnitude $\Delta$ peaks sharply at $\theta=\pi/2$ and decreases rapidly with deviation from this direction, as depicted in Fig. \ref{fig2}(d). This strongly angle-dependent edge gap mechanism naturally explains the anti-corner states observed in circular nanoflake: edge-localized modes persist along most boundaries but vanish sharply along specific corners, giving rise to the anti-HOT phase.

With the above phenomenology established, we now turn to its microscopic origin. The key lies in how each layer’s topology is determined by the distribution of topological charges within the BIS \cite{zhang_dynamical_2018,zhang_dynamical_2019}. The BIS is defined as the momentum contour where the mass term of the Hamiltonian vanishes $d_z^\alpha\left(\mathbf{k}\right)=0$. Here, topological charges are pinned at the four high-symmetry momenta points $\Gamma=\left(0,0\right), X=\left(\pi,0\right), Y=\left(0,\pi\right)$ and $M=\left(\pi,\pi\right)$, as indicated by four green dots in Fig. \ref{fig2}(e). Recent studies have shown that Chern phases can be further classified by labeling the high-symmetry point enclosed by the BIS \cite{wan_classification_2025}. In this picture, the Chern number determines the chirality, while the BIS-enclosed symmetry point dictates where the edge states emerge in momentum space. In our system, layer $1$ has a BIS enclosing the $\Gamma$ point [see Fig. \ref{fig2}(e)], giving $C_{1\Gamma}=+1$, whereas layer $2$ encloses the X point [see Fig. \ref{fig2}(f)], leading to $C_{2X}=-1$. Since the edge momentum is fixed by the enclosed topological charge, the edge states project to different momenta depending on the edge orientation. Along the $\theta=\pi/2$ edge, both topological charges project to $k_y=0$, allowing interlayer hybridization and opening a gap. In contrast, along the $\theta=0$ edge, the topological charges project separately to $k_x=0$ and $k_x=\pi$, so the corresponding edge states remain well separated and gapless. For all other edge angles, the projected momenta do not match, preventing significant interlayer hybridization. As a result, the band gap magnitude exhibits a strongly angle-dependent behavior: peaking sharply at $\theta=\pi/2$ and decreasing rapidly as the edge angle deviates from this direction. In short, the realization of the anti-HOT phase requires that the BIS enclose topological charges located at distinct high-symmetry points, with these charges carrying opposite Chern numbers. This condition enables interlayer hybridization only along specific boundary directions where the topological charges project to the same momentum space. The angle-dependent edge hybridization leads to the formation of anti-corner states, which is the hallmark of the anti-HOT phase.

\begin{figure}
	\centering
	\includegraphics[width=\columnwidth,angle=0]{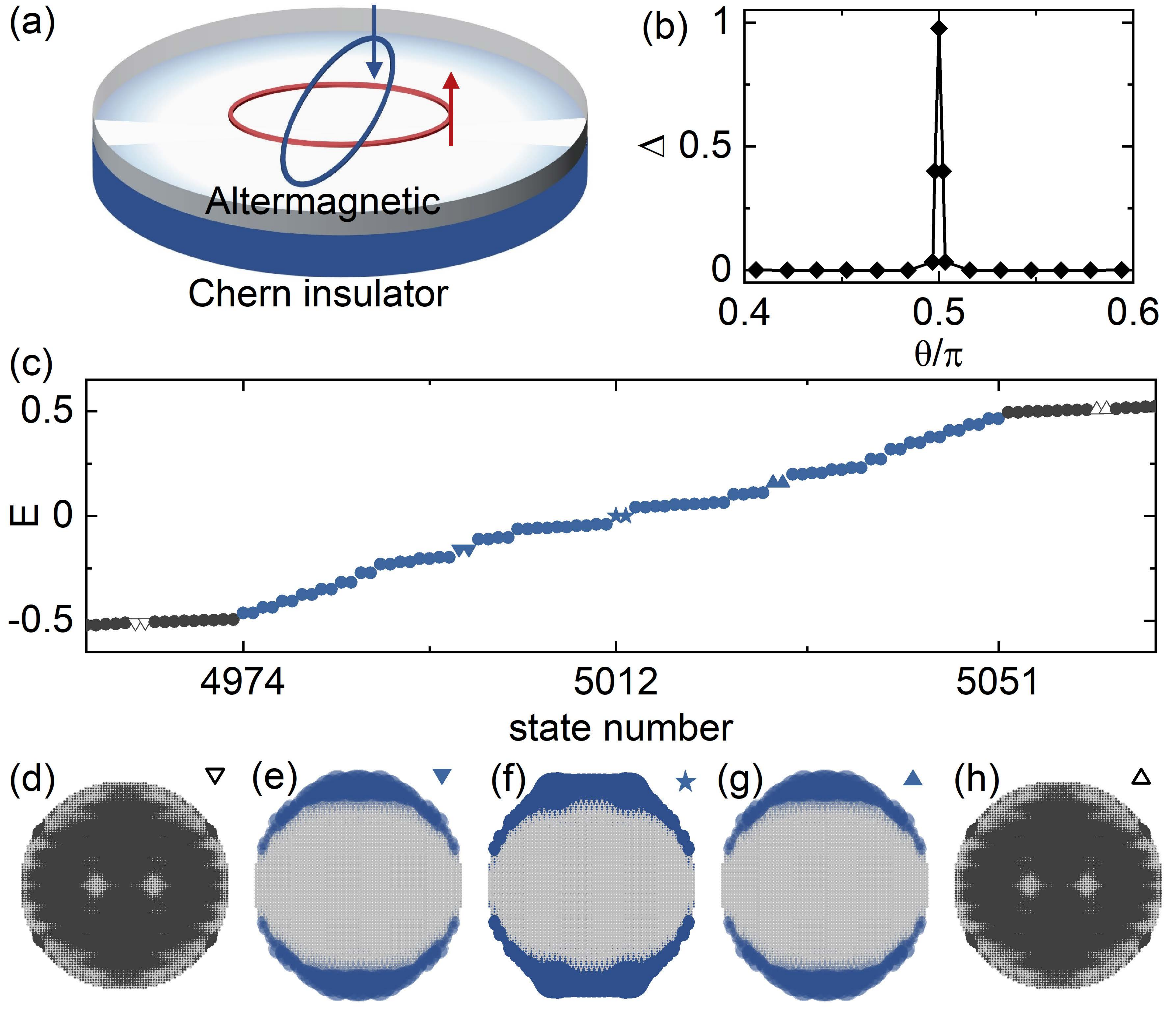}
	\caption{(a) Schematic illustration of a Chern insulator with $d$-wave altermagnetic coupling. (b) Energy gap $\Delta$ vs edge angle $\theta$ with width $L=300$. (c) Spectrum of a circular nanoflake, highlighting anti-corner (blue) and bulk (black) states. (d-h) Spatial distributions of the representative states marked in (c). Parameters: $R=40, m=B=1, J=0.9$.}
	\label{fig3}
\end{figure}

\emph{Anti-HOT phase in altermagnetic system}.---The bilayer Chern insulator model above identifies the essential conditions and underlying mechanism for realizing the anti-HOT phase. Guided by this insight, we consider altermagnets \cite{smejkal_emerging_2022,krempasky_altermagnetic_2024,ghorashi_altermagnetic_2024,liu_absence_2024,zhou_manipulation_2025,chen_unconventional_2025,yi_spin_2025,wan_altermagnetisminduced_2025,wan_helical_2025,wan_interplay_2025}, whose intrinsic momentum-space spin anisotropy naturally produces BIS enclosing distinct high-symmetry points, making altermagnetically coupled Chern insulators a promising platform for the anti-HOT phase.
Motivated by this symmetry property, we introduce $d$-wave altermagnetic coupling into a monolayer Chern insulator to construct a lattice model for the anti-HOT phase, as schematically shown in Fig. \ref{fig3}(a). The Hamiltonian of the coupled system reads:
\vspace{-0.5em}
\begin{align}
H=&[m-2B\left(2-\cos{k_x-\cos{k_y}}\right)+2J\left(\cos{k_y}-\cos{k_x}\right)]\sigma_z\nonumber \\
&+\sin{k_x}\sigma_x+\sin{k_y}\sigma_y,
\label{eq2}
\end{align}
where $m$ is the mass parameter, $B$ and $J$ denote the hopping and altermagnetic strengths, respectively. In Fig. \ref{fig3}(b), in the nanoribbon geometry, the energy gap reveals a sharp maximum at $\theta=\pi/2$, indicating strong angle dependence. In the nanoflake geometry, the near-zero-energy spectrum [Fig. \ref{fig3}(c)] separates trivial bulk states (black dots) from a series of anti-corner states (blue dots). To illustrate their spatial distribution, we show three representative anti-corner states with blue downward triangles, pentagrams, and upward triangles in Fig. \ref{fig3}(c). These states extend along most boundaries but vanish sharply at specific corners [see Figs. \ref{fig3}(e)-\ref{fig3}(g)], which constitutes the key fingerprint of the anti-HOT phase. While the bulk states [Figs. \ref{fig3}(d) and \ref{fig3}(h), marked by white downward triangle and upward triangle] show an extended spatial distribution across the nanoflake. Nanoribbon calculations for the altermagnetic system further confirm the angle-dependent behavior (see Fig. S3 and Sec. S5 of the Supplemental Material \cite{supplemental}). Together, these results demonstrate that the direction-dependent energy gaps arise naturally from the altermagnetic coupling, providing a clear realization of the anti-HOT insulator.

\begin{figure}
	\centering
	\includegraphics[width=\columnwidth,angle=0]{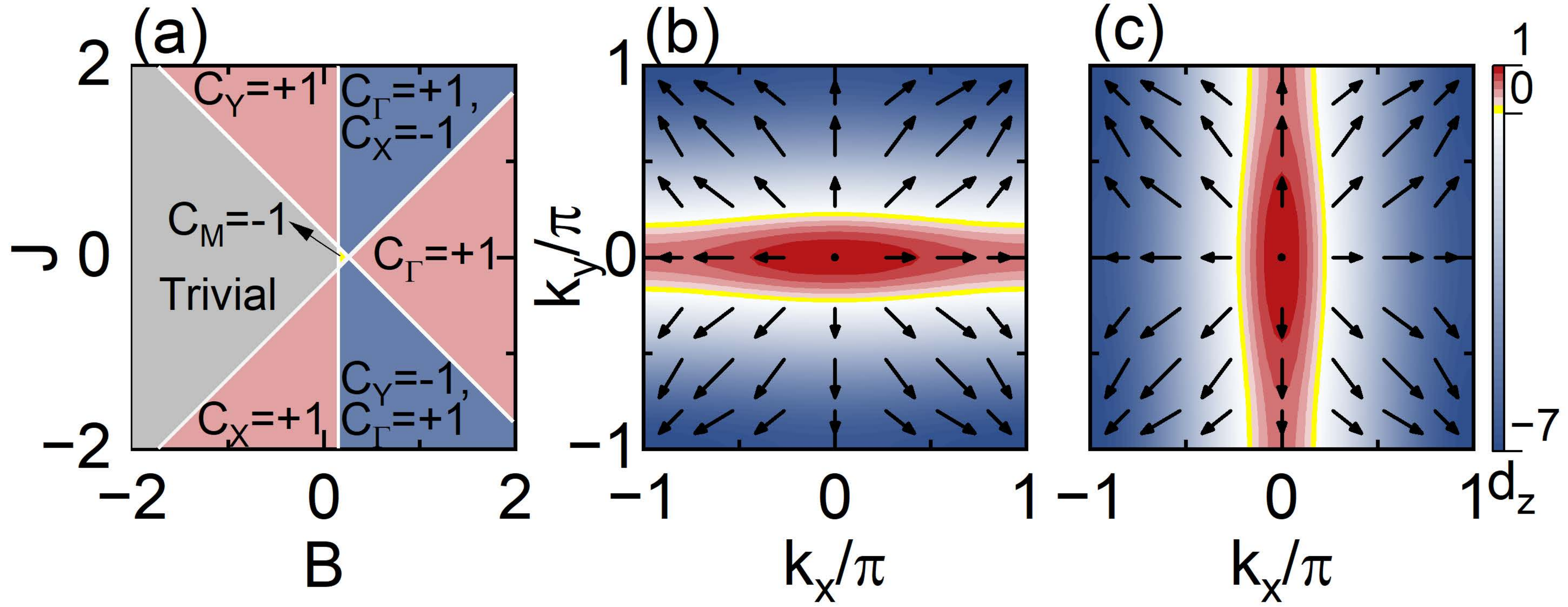}
	\caption{(a) Phase diagram in the ($B, J$) plane with $m=1$. Gray, red (yellow), and blue regions denote trivial, first-order topological, and anti-HOT phases, where the BIS encloses 0, 1, and 2 high-symmetry points, respectively. (b, c) Spin textures for two anti-HOT phase. Arrows (colors) indicate in-plane (out-of-plane) components. Parameters: $B=1,J=0.9$ in (b) and $B=1,J=-0.9$ in (c).}
	\label{fig4}
\end{figure}

By tuning the Chern-insulator hopping parameter $B$ and the altermagnetic coupling strength $J$, we obtain the phase diagram shown in Fig. \ref{fig4}(a). The boundaries between distinct topological phases can be determined analytically, that is, by solving the condition where the bulk energy gap closes at zero. For the Hamiltonian considered, the bulk gap closes under four conditions: $m=0, J=B-m/4, J=-B+m/4$, and $B=m/8$. In our study, we fix $m=1$, leaving three lines $J=B-1/4, J=-B+1/4$, and $B=1/8$ in the $J$-$B$ plane, as indicated by the white lines in Fig. \ref{fig4}(a). These three lines divide the phase diagram into seven distinct regions, each corresponding to a unique configuration of BIS-enclosed topological charges. In the gray region ($J>B-1/4$, $J<-B+1/4$, $B<1/8$), the BIS does not enclose any topological charges, corresponding to a trivial phase. When the BIS encloses a single topological charge with positive chirality, the system enters a first-order topological phase with Chern number $+1$. These phases are labeled as $C_\Gamma=+1$ ($J<B-1/4$, $J>-B+1/4$), ${C}_X=+1$ ($J<B-1/4$, $B<1/8$), and $C_Y=+1$ ($J>-B+1/4$, $B<1/8$), depending on which high-symmetry point is enclosed [see red regions in Fig. \ref{fig4}(a)]. In the small yellow region ($J>B-1/4$, $J<-B+1/4$, $B>1/8$), the BIS encloses the topological charge at the high-symmetry point $M$ with negative chirality, representing a conventional first-order topological phase $C_M=-1$. Most notably, the two blue regions correspond to the anti-HOT phase. In these regions, the BIS encloses two topological charges with opposite Chern numbers at distinct high-symmetry points, resulting in an overall Chern number of zero, which is the hallmark of the anti-HOT phase. Specifically, the enclosed topological charges are located at the high-symmetry points $\Gamma$ and $X$ ($J>B-1/4$, $J>-B+1/4$, $B>1/8$), or located at the high-symmetry points $\Gamma$ and $Y$ ($J<B-1/4$, $J<-B+1/4$, $B>1/8$). Spin-texture analysis reveals that the enclosed topological charges at distinct high-symmetry points carry opposite chiralities [see Figs. \ref{fig4}(b) and \ref{fig4}(c)], leading to two sets of counter-propagating edge states at different momenta. For a given edge orientation, the projections of these edge states either overlap, resulting in selective hybridization and the opening of a gap along specific edges, or remain separated and gapless, in agreement with the toy model mechanism.
The comparison between Figs. \ref{fig4}(b) and \ref{fig4}(c) highlights the tunability of this effect: in the former case, edge states disappear around $\theta=\pi/2$, while in the latter they vanish along $\theta=0$.
Figure S4 and section S6 of the Supplemental Material \cite{supplemental} confirms these conclusions in finite circular nanoflakes. Altogether, the phase diagram demonstrates that the anti-HOT phase is stabilized by topological charges of opposite chirality enclosed by the BIS. The angle at which edge states vanish is determined by the specific topological charges enclosed. This establishes both the robustness and directional tunability of the anti-HOT phase, offering a clear guideline for experimental realization.

\emph{Topological classification of anti-HOT phases}.---To classify and characterize the anti-HOT phases identified above, we analyze the distribution of topological charges in our model. All topological charges are pinned to four high-symmetry points $\Gamma,X,Y$, and $M$. This originates from inversion symmetry: the two-dimensional inversion operator can be written as $P=\hat{P}\otimes \hat{R}_{2D}$, where $\hat{P}=\sigma_z$ acts on the spin subspace and $\hat{R}_{2D}$ inverts the spatial coordinate $\mathbf{R}\rightarrow-\mathbf{R}$.
For a system preserving inversion symmetry, $\hat{P}H(\mathbf{k})P^{-1}=H(-\mathbf{k})$.
At inversion-invariant momenta $\Lambda_i\in \left\{\Gamma,M,X,Y \right\}$, this reduces to $\hat{P}H(\Lambda_i)P^{-1}=H(\Lambda_i)$. Since $H\left(\Lambda_i\right)=d\left(\Lambda_i\right)\cdot \hat{\sigma}$, one obtains $d_x\left(\Lambda_i\right)=d_y\left(\Lambda_i\right)=0$, ensuring that topological charges must reside at these high-symmetry points. Thus, inversion symmetry provides the fundamental reason why all topological charges in our system are pinned at symmetry points.

Building on this observation, we define the BIS-enclosed topological charge set
$\mathcal{C}_{BIS}=\left\{\mathcal{C}_\beta\mid\beta\in{\Gamma,X,Y,M},\beta\mathrm{\ enclosed\ by\ the\ BIS\ } \right\},$
where $\mathcal{C}_\beta$ is the topological charge associated with the high-symmetry point $\beta$. Only the enclosed topological charges are relevant for the low-energy boundary physics, since they determine the momenta at which edge states appear. A phase in which the BIS encloses a single topological charge corresponds to a conventional Chern state, while the anti-HOT phase arises when two inequivalent points with opposite topological charges are simultaneously enclosed. In this situation, two sets of counter-propagating edge states appear at distinct projected momenta, leading to selective hybridization and gapping along certain edge orientations while leaving others gapless. Crucially, phases with different $\mathcal{C}_{BIS}$ cannot be adiabatically connected, since changing the enclosed set requires the BIS to cross a topological charge, which necessarily closes the bulk gap and drives a topological transition.
The BIS-enclosed topological charge set therefore constitutes a genuine topological invariant for classifying anti-HOT phases. It provides a symmetry-based foundation for the numerical and analytical results presented above and directly explains the phase diagram in Fig. \ref{fig4}(a), where different regions are distinguished by the topological charges enclosed and their edge projections.

\emph{Conclusion}---
In summary, we have introduced anti-HOT insulators as the dual counterpart of conventional HOT insulators. Unlike HOT insulators where topological states localize at corners, anti-HOT insulators exhibit anti-corner states that vanish at specific corners. We present its topological origin, realization platform, topological invariant, and key properties. Our work establishes anti-topology as a new theoretical framework, offering a fresh perspective on topological phenomena.

\emph{Acknowledgements}.---This work was financially supported by the National Key R and D Program of China (Grant No. 2024YFA1409002), the National Natural Science Foundation of China (Grants Nos. 12447147, 124B2069, 12374034 and 12074097), the Quantum Science and Technology-National Science and Technology Major Project (Grant No. 2021ZD0302403), the China Postdoctoral Science Foundation (Grant No. 2024M760070), and Natural Science Foundation of Hebei Province (Grant No. A2024205025). We also acknowledge the High-performance Computing Platform of Peking University for providing computational resources.

\emph{Code availability}.---The code that supports the findings of this study is available from the corresponding author upon reasonable request.

\emph{Author contributions}.---Q.-F. S. conceived the work and designed the research strategy. C.-M. M. and Y.-H. W. carried out the analytical analysis and numerical calculations under the supervision of Q.-F. S. All authors wrote the paper together.

\bibliography{AHOTI}

%\vspace{1em}
%\vspace{1em}

\end{document}